\documentclass[10pt]{article}
\usepackage{graphicx}
\usepackage{pdflscape}

\usepackage{epsfig}
\usepackage{latexsym}
\usepackage{amsmath}
\begin{document}

\title{\Large \bf   Toward fits to scaling-like data,\\ but with inflection points \\\& generalized Lavalette function }

\author{ \large \bf  Marcel Ausloos$^{1,2,3}$$ $\\ \\$^1$  7B La Roche Pertuade, \\544 chemin du Romanet, F-84490, St. Saturnin-les-Apt, France
 \\ $^\#$email:  ausloosm@fastmail.fm$ $\\   $^2$ e-Humanities, \\Royal Netherlands Academy of Arts and Sciences, \\  Joan Muyskenweg 25, 1096 CJ Amsterdam, The Netherlands\\ $^\#$email:  marcel.ausloos@ehumanities.knaw.nl $ $\\ $^3$  R\' es. Beauvallon, rue de la Belle Jardini\`ere, 483/0021\\
B-4031, Li\`ege Angleur, Euroland \\$^*$email: marcel.ausloos@ulg.ac.be\\ }

\maketitle

 \begin{abstract}

Experimental and empirical data are often analyzed on log-log plots in order to find some scaling argument for the observed/examined phenomenon at hands, in particular for rank-size rule research, but also in  critical phenomena in thermodynamics, and in fractal geometry. The fit to a straight line on such plots is not always  satisfactory. Deviations occur at low, intermediate and high regimes along the log($x$)-axis. Several improvements of the mere power law fit are discussed, in particular through a Mandelbrot trick at low rank and a Lavalette power law cut-off at high rank. In so doing, the number of free parameters increases. Their meaning is discussed, up to the  5 parameter free super-generalized  Lavalette law and the 7-parameter free hyper-generalized Lavalette law. It is emphasized that the interest of the  basic 2-parameter free Lavalette law and the subsequent generalizations resides in its "noid" (or sigmoid, depending on the sign of the exponents) form on a semi-log plot; something incapable to be found in other empirical law, like the Zipf-Pareto-Mandelbrot law. It remained for completeness to invent a simple law showing an inflection point on a \underline{log-log plot}. Such a law can  result from a transformation of the Lavalette law through $x$ $\rightarrow$  log($x$), but this meaning is theoretically unclear. However, a simple linear combination of two basic Lavalette law is shown to provide the requested feature. Generalizations taking into account two super-generalized or hyper-generalized Lavalette laws are suggested, but need to be fully considered at fit time on appropriate data.

\end{abstract}

\vskip 0.5truecm
  Keywords :  graphs, plots, nonlinear laws
\vskip 0.5truecm

\vskip 0.5truecm
\section{  Introduction}\label{sec:introduction}
In recent years, following the rise in the understanding of critical phase transitions \cite{HESbook1} through the notion of critical exponents, many results have been presented on log-log graphs.  It should be emphasized at once that the search for a straight line fit  on such a graph is of interest  when the hypothesis of  scaling  is appropriate for the examined property or effect.  Then, the slope on the plot gives  some indication of some characteristic exponent  at the phase transition because the underlying analytical  function, the excess free energy  \cite{HESbook1},  has a homogeneity property.  Two other major scientific  concepts, related to some underlying scaling hypothesis,  have also led to examining log-log plots for various quantities: one is the notion of fractal dimension \cite{Mandelbrot}, the other is the rank-size relationship through so called Zipf plots \cite{z1}.  

\begin{figure}
  \centering
  \includegraphics  [height=14cm,width=14cm] 
  {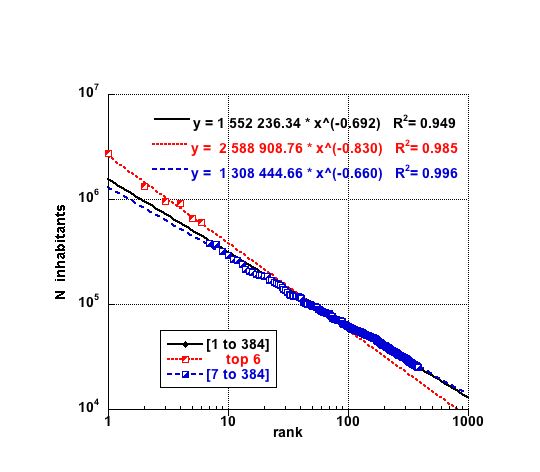}
\caption{  The 384 largest Italian cities ranked by decreasing order  of  
their population size, pointing
to a drop  after the main 6;    different power law fits  for the whole range  (black line) or   when distinguishing two regimes  (red and blue line) are  indicated  with their corresponding correlation coefficient $R^2$ }\label{fig6a:plotbigITcitieslolo}
\end{figure}

It is  often discussed whether the scaling law should hold over many decades of the $x$-axis variable, -whatever the $x$-axis (reduced temperature $\epsilon$, bin size $n$, rank $r$, ...).  Officially, this "many decades  validity" should be the case, if a scaling law fully holds. However,  phenomena for which (quasi) straight lines are seen on a log-log plot are  rarely found, -  outside laboratories or computer simulations. Yet, there is no harm in recognizing that  such a straight line existing on a small  $x$-axis  range indicates the presence of a specific regime; see for example the case of the population size of large italian cities, as illustrated in Fig. \ref{fig6a:plotbigITcitieslolo}, for which two regimes rather than a single one can be imagined. Therefore, weak scaling can be accepted as physically suggestive within finite $x$-axis ranges. 

Nevertheless, the data can often present convex or concave shapes, and often gaps, jumps, drops  (see Fig. \ref{fig6a:plotbigITcitieslolo}) or shoulders.   Such a large variety of basic shapes  demands to pursue some systematic inquiry of the simplest appropriate analytical forms representing complicated data. Much difficulty resides in (interpreting and)  theoretically manipulating  inflection points, - often visible when a line is drawn through the data "for the eye".

The  2-parameter free power law (using thereafter the  discrete variable $r$ for the $x$-axis)
\begin{equation}\label{Zipfeq} 
y_r = \cfrac{a}{r^\alpha},
\end{equation}
on a log-log plot is referred to Zipf's plot. 
 Zipf had  thought that   the
  particular case  $\alpha=1$ 
represents a desirable situation, in which   forces of concentration
balance those of decentralization \cite{z1,gabaix}. Such a case is called  the rank-size $rule$ \cite{gabaix}-\cite{PhA391.12.767ranksizescaling}. 
 Thus the scaling exponent  $\alpha $ can be used to judge whether or not the size distribution is close 
to  some optimum  (equilibrium) state.

The pure power-law distribution, for a continuous variable,
reads
\begin{equation}\label{eq0} p(k) = \frac{k^{-\gamma}}{\zeta(\gamma)}
   \end{equation} 
where 
  $k$ is a positive integer usually measuring some variable
of interest; 
 $p(k)$ is the probability of observing the value $k$;
  $\gamma$  is the power-law exponent;
and $\zeta(\gamma)$ $\equiv$$ \sum_{k=1}^{\infty} k^{-\gamma} $ is the Riemann zeta function; note that $\gamma$, in Eq.(\ref{eq0}) must be greater than 1 for the Riemann zeta function to be finite.

However,  the fit to a straight line on a log-log plot is not always truly perfect, as any reader has surely had the experience considering various data with expected scaling.  The error bar (e.g., on $\gamma$)   can be very large for a $R¬2$ or $\chi^2$ test point of view. Moreover, broadly used
methods for fitting to the power-law distribution 
 provide biased estimates for the power-law exponent \cite{0402322fittingtopowerlaw_v3}. 
 
 The deviations occur in various regimes along the log($x$)-axis.  

\begin{figure}
\centering
 \includegraphics [height=8.0cm,width=11.0cm]{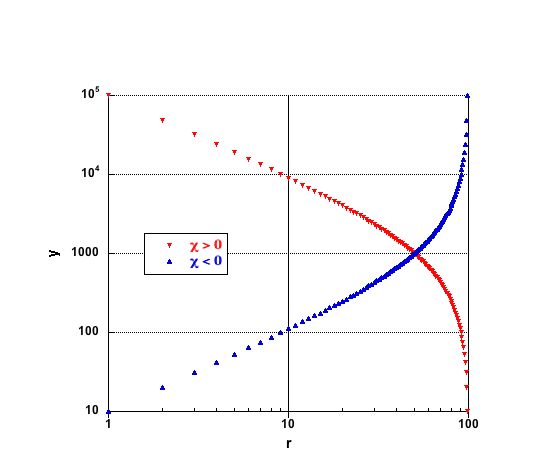}
\caption   {   Lavalette function, Eq.(\ref{Lavrevsigmoidal}) with either $\chi >0$ (red dots) or $<0$ (blue dots) on a log-log plot, for $N$ = 100 and $\hat{\kappa}=10^6$ 
  } 
 \label{fig1Lavposneglolo} 
 \end{figure}

\begin{figure}
\centering
 \includegraphics [height=8.0cm,width=11.0cm]{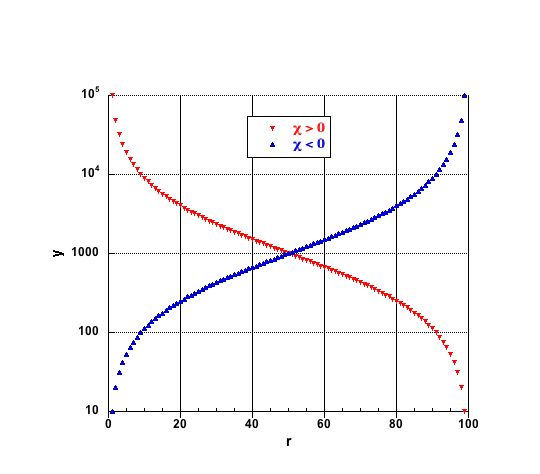}
\caption   {  Lavalette function, Eq.(\ref{Lavrevsigmoidal})  with either $\chi >0$ (red dots) or $<0$ (blue dots) on a  semi-log plot, for $N$ = 100 and $\hat{\kappa}=10^6$, emphasizing the inflection points at $r=N/2$ 
  } 
 \label{fig2Lavposneglilo} 
 \end{figure}

 When  the data crushes at high $x$-axis value, Lavalette suggested
\cite{Lavalette} to use the 2-parameter free  ($\kappa$, $\chi$) form
 \begin{equation} \label{Lavalette2} 
y(r)= \kappa\; \big[\frac{N\;r}{ N-r+1} \big] ^{-\chi}  
\end{equation} 
in which  the role of $r$ as the independent variable, in Eq.(\ref{Zipfeq}), is taken by the ratio $r/(N - r + 1)$ between the descending and the ascending ranking numbers;  $N$ is the number of data points on the $x$-axis,  and $\chi\ge0$; the +1 role in  $(N - r + 1)$  is easily understood.  Other ways of writing  this 2-parameter Lavalette  form  function are of interest 
\begin{eqnarray} \label{Lavrevsigmoidal}
y(r)=\kappa\; (N\;r/(N-r+1))^{-\chi} 
 \; \equiv \;  \kappa\;  N^{-\chi}\; (r/(N-r+1))^{-\chi} 
 \\   \equiv \; 
 \kappa\;  (N\;r)^{-\chi}\;(N-r+1)^{+\chi}\; \\
 \equiv  \hat{\kappa}\;   r^{-\chi}\;(N-r+1)^{+\chi}\;
. \end{eqnarray}
in order to be emphasizing a \underline{power law decay} with  a \underline{power law cut-off}.
The interest in such a function which is  strictly decreasing,  Fig.\ref{fig1Lavposneglolo}, from infinity at $r=0$ under a $r^{-\chi}$ law  to a zero value at $r=N+1$ as $(N-r+1)^{+\chi}$,  best appears on a  semi-log plot, Fig. \ref{fig2Lavposneglilo}:  observe the inflection point presence at $r=N/2$. The slope $s$  at such a point  is equal to $-4\chi\; \frac {N+1}{N(N+2)}$ which for "large r" $\sim -4\chi (1/N)(1-1/N)$. In some sense, it is realistic to reproduce this intermediary regime as $y \sim e^{-sr}$.

 When $\chi \le 0$,  - not a rank-size rule case, the function is increasing, - it is a
  flipped Lavalette function.   Both functions, i.e. with $\chi\ge0$ or $\chi \le 0$,  are shown  in Fig. \ref{fig1Lavposneglolo} on  a log-log plot, - where the shape is apparently simple, i.e. a power law followed by a  sharp cut-off indeed,  and   on a semi-log plot in Fig. \ref{fig2Lavposneglilo}, where the shape is "more trivial".   On a semi-log plot, Eq.(\ref {Lavalette2}) with $\chi\le0$, gives  a flat   N-shape "$noid$" function   (which could be called a "reverse sigmoidal")    near its inflection point, which with the correspondingly flat  S-shape, but  nevertheless called "$sigmoid$" function, allows to cover  various convex and concave data display shapes\footnote{Recall that these functions/shapes are found in laboratory when measuring the (I,V) characteristics of   junctions or diodes; they present an N or S shape, beside the Ohm law.  The sigmoid or noid form are also describing speculator's different strategies on the stock market \cite {MA359kinetic} .}.

   \begin{figure}
\centering
 \includegraphics [height=8.0cm,width=11.0cm]{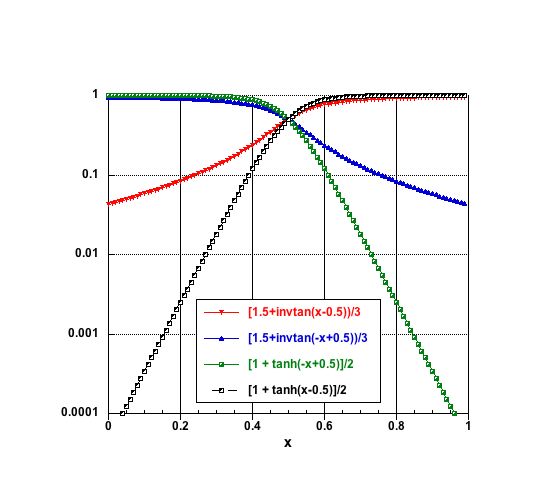}
\caption   {  Display of types of sigmoid functions ($invtan(x)$ and $ tanh(x)$) on semi-log axes
} 
 \label{Plot10sigmoidCDEFlilo} 
 \end{figure}

    \begin{figure}
\centering
 \includegraphics [height=8.0cm,width=11.0cm]{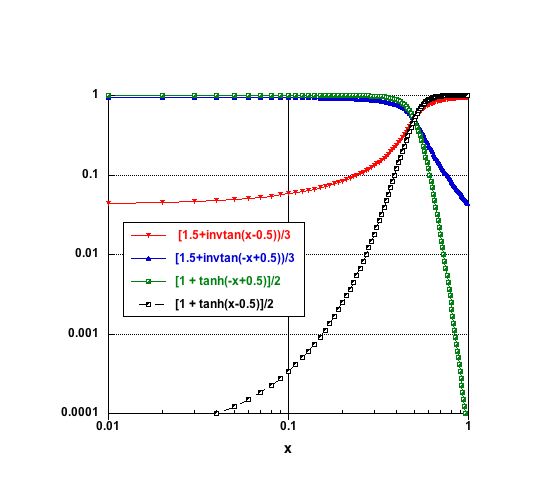}
\caption   {   Display of types  of sigmoid functions ($invtan(x)$ and $tanh(x)$) on log-log axes
 } 
 \label{Plot10sigmoidCDEFlolo} 
 \end{figure}
 
  No need to recall that   other   often seen (or used)  2-parameter free  (amplitude and slope at inflection point) have a sigmoid  shape; they are $tanh (\gamma x)$  and 
and $invtan(\gamma x)$. There is of course no need to represent such well known functions \underline{on classical graphs}. They are rarely seen, thus shown on \underline{semi-log} and \underline{log-log} plots in Figs. \ref{Plot10sigmoidCDEFlilo}-\ref{Plot10sigmoidCDEFlolo}  respectively. The functions have been adapted and scaled in order to read them on appropriate graphs, for comparison with other functions\footnote{It should be obvious to the reader that all these S or N shape functions can occur on different types of plots.  The question is whether it can be trivially made $x \rightarrow log(x)$,    whether this "transformation"  has  any impact on data analysis, and whether some  theoretical hypothesis can sustain/justify such a transformation.}.

\begin{figure}
\centering
\includegraphics  [height=14cm,width=14cm] 
 {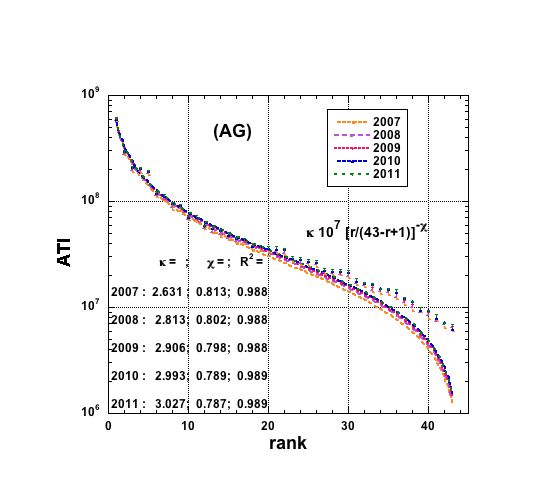}
\caption   { Basic 2-parameter free fit Lavalette law to the Aggregated Income Tax (ATI) of  the $N$=43  cities, ranked in decreasing ATI order,  in the province of Agrigento, IT, for recent years. Note the high regression coefficient, but not the visually pleasing fit at high rank  ($r\ge22$)}   \label{fig:ITAG}
\end{figure}

   A technical point is in order here.
  Note that $N$  (as a factor of $r$, e.g. in Eqs.(\ref{Lavalette2}-\ref{Lavrevsigmoidal}) is not really needed. In fact it is more usefully  replaced, at fit time,  by some simple factor having the order of magnitude of $y(N/2)$.  This was  made in Fig.\ref{fig:ITAG}, for example. The Aggregated Income Tax of the  43 cities in the province of Agrigento (AG) in Italy was ranked in decreasing order, for each available year  in [2007-2011], from the Italian Ministery of Economy,  and fitted by an adapted simple Lavalette law, i.e. $\kappa\; 10^7\; \big[r/43-r+1\big]^{-\chi}$. Note the high regression coefficient values, but  a not so visually pleasing fit at high rank

 Finally, considering cut-offs at high rank, there is on the contrary  not much discussion in the literature on the wide flattening of the data at high rank, - although such cases are encountered,  e.g.  in co-author ranking  \cite{Sofia3,MABSS,[HB],[JM],[GR]}, and in other  "very long flat tail" cases. 

For completeness, other 2-parameter free simple functions are recalled in Sect. \ref{sec:afewformulae}.

    \begin{figure}
\centering
 \includegraphics [height=8.0cm,width=11.0cm]{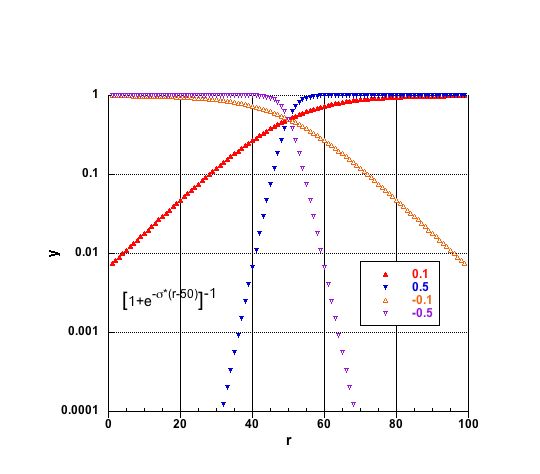}
\caption   {  Logistic function, Eq.(\ref {Vlog3}),   on semi-log axes
 } 
 \label{Fig3sigmoidslilo} 
 \end{figure}
 
     \begin{figure}
\centering
 \includegraphics [height=8.0cm,width=11.0cm]{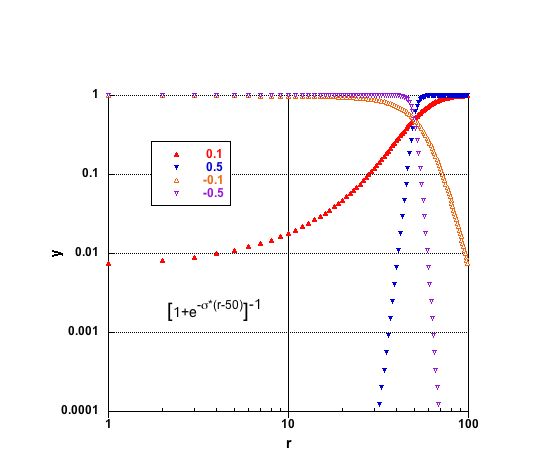}
\caption   {  Logistic function, Eq.(\ref {Vlog3}),  on log-log axes
 } 
 \label{Fig4sigmoidslolo} 
 \end{figure}
 
\section{A few 3-parameter free functions}\label{sec:3free}
Having, introduced well known  2-parameter free functions, to represent complicated  data, let us turn on functions with 3 (or more, see  below) free parameters,  toward elaborating an attempt on how to take into account deviations from   simple data approximations by power law-like lines.

\subsection{Logistic or Verhulst function}\label{sec:Vlog3}

For completeness, recall that the 3-parameter  ($\sigma$,  $y_M$,  and $r_{M/2}$ )  sigmoid  forms are well represented through the usually called 
  Verhulst logisitic  \cite{Vlog3} 
   \begin{equation} \label {Vlog3}
y(r)=   \frac{y_M}{1+  e^{-\sigma^*\; (r-r_{M/2})}}.
\end{equation}
based on the exponential  (growth) function, but invented for  limiting the maximum value such a growth function can reach. This well known function does not need to be shown on   an ordinary scale  graph.  The function is topologically similar to $tanh (\gamma x)$ 
and $invtan(\gamma x)$. However, 
it  is unusual to  see this sigmoid function represented on a log-log plot or on a semi-log plots;  whence  this is shown, in Fig. \ref{Fig3sigmoidslilo} and Fig.\ref{Fig4sigmoidslolo}, for different $\sigma^*$ values (with $r_M$=100), pointing to non trivial shapes, - also different from those on Figs. \ref  {Plot10sigmoidCDEFlilo}-\ref {Plot10sigmoidCDEFlolo}, as the reader can usefully observe by himself..

Interestingly, and "obviously", it can noted that  some data  which could be represented by the Verhulst logisitic, Eq.(\ref{Vlog3}), can be  transformed  through a simple combination, $ [y(r)/( y_M-y(r))] $, into some $Y(r)$ which is   $\equiv e^{-\sigma^*\; (r-r_{M/2})}$. Therefore a semi-log plot of  $Y(r)$ vs.  $r$ expectedly leads to a graph with a straight slope from which parameters can be easily deduced \cite{PNAS75.78.4633-7-Montroll-socialforces}; practically, $y_M$ can be used as an appropriate input parameter to optimize the fit.

\subsection{Zipf-Mandelbrot function}\label{sec:ZMP3}
When the data upsurges at low rank ($r\sim1$), on a log-log plot,  as in  \cite{jefferson1989geography}, one   mentions a "king effect" \cite{Stx3}, apparently first emphasized in city population size distributions \cite{jefferson1989geography}.  When the data flattens, below the expected straight line, at low $r$ values,  when a so  called "queen effect" occurs \cite{Sofia3}, it is best to modify Eq.(\ref{Zipfeq})  into
a  3-parameter free form, called the Zipf-Mandelbrot-Pareto  (ZMP3) law  \cite{FAIRTHORNE},  which reads
 \begin{equation} \label{ZMeq3}
y(r)=\hat{c}/(\eta+r)^{\zeta} \;\equiv \; [c/(\eta+r)]^{\zeta},
\end{equation}
since obviously $y(0)$ takes a finite value. The value $\eta$ is understood as a measure of the "harem" \cite{MABSS}, - as seen in co-authors of papers distributions.

\subsection{Generalized 2-exponent Lavalette function}\label{sec:Lav3}

There is no reason for which the behavior near the crushing point be of  (analytically) identical  type as the vertical  asymptotic behavior at low rank. 
The basic  2-parameter Lavalette  form (Lav3) 
Eq.(\ref{Lavalette2}) 
can be generalized 
as   a
3-parameter  free form   
\begin{itemize} \item
 e.g. allowing two exponents ($\chi$ and $\xi $) \cite{JoI1.07.155Mansilla}:
  \begin{equation} \label{Lavalette3b}
 \;\;  y_N(r)= \kappa\;  \frac{(N\;r)^{-\chi}}  { (N-r+1)^{-\xi}  }
\end{equation}
which is  emphasizing the number of data points as in Eq.(\ref{Lavalette2}), but can be  simply written

\begin{eqnarray}\label{Lavalette3}
\;\;  y(r)=   \Lambda \frac{\big[ r\big]^{-\phi}} { \big[N-r+1\big] ^{-\psi} }\; \equiv \;    \Lambda\;  \big[ r\big]^{-\phi}\; \big[N+1-r\big] ^{+\psi}  
 \\   \equiv \; 
\Lambda\; (N+1)^{\psi-\phi} \big[ \frac{r}{N+1}\big]^{-\phi}\; \big[1-\frac{r}{N+1}\big] ^{+\psi}   \\
 \equiv  \hat{\Lambda}\;   u^{-\phi}\;(1-u)^{+\psi}\;
\end{eqnarray}
\end{itemize}
In fact, the case $\phi >0$ and $\psi <0$  is the Feller-Pareto function. The case   $\phi =-1$ with  $\psi =+1$ is the Verhulst   function introduced in the  right hand side of the (logistic) evolution differential equation. For $\xi=0$ or $\psi=0$, it  has the Eqs.(\ref{Zipfeq})-(\ref{eq0}) form.

\begin{figure}
\centering
 \includegraphics [height=7.4cm,width=11.0cm]{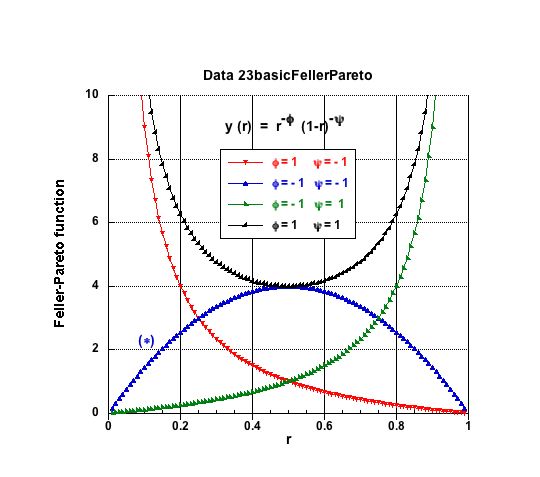}
\caption   {  Feller-Pareto function,  $y(r) = r^{-\phi}   (1 - r)^{-\psi}$,  but extended to allow different signs (and possible vlues) for    $\phi$ and $\psi$; for readability the  amplitude of the  $\phi=-1$ and $\psi=+1$ case has been multiplied by a   factor 16 as pointed out  by (*)} 
 \label{Plot23basicFellParlili} 
 \end{figure}

\begin{figure}
\centering
 \includegraphics [height=7.4cm,width=11.0cm]{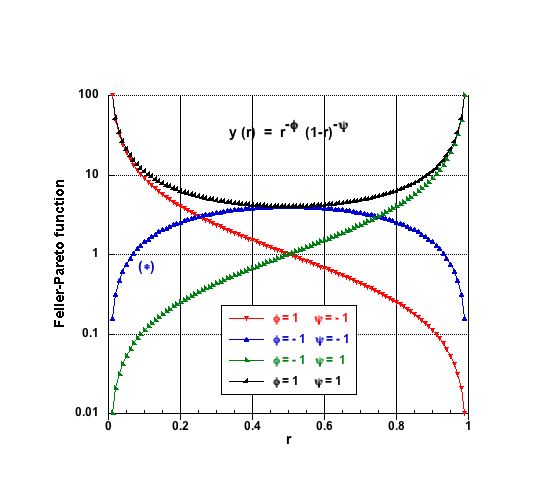}
\caption   { Feller-Pareto function,  $y(r) = r^{-\phi}   (1 - r)^{-\psi}$,  on a semi-log plot, but extended to allow different signs (and possible values) for    $\phi$ and $\psi$; for readability the  amplitude of the  $\phi=-1$ and $\psi=+1$ case has been multiplied by a   factor 16 as pointed out  by (*)}
 \label{Plot23basicFellParlilo} 
 \end{figure}

\begin{figure}
\centering
 \includegraphics [height=7.4cm,width=11.0cm]{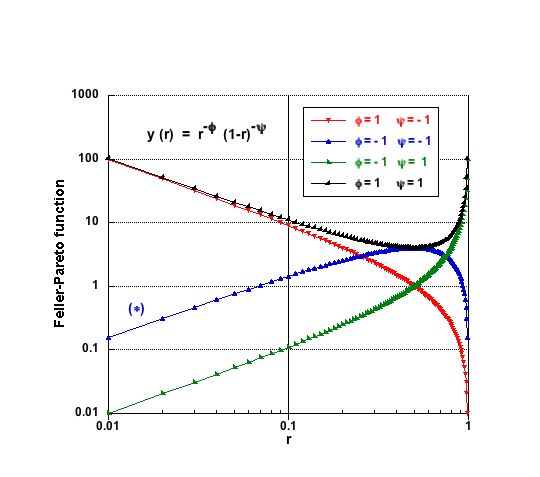}
\caption   { Feller-Pareto function,  $y(r) = r^{-\phi}   (1 - r)^{-\psi}$,  on a slog-log plot, but extended to allow different signs (and possible values) for    $\phi$ and $\psi$; for readability the  amplitude of the  $\phi=-1$ and $\psi=+1$ case has been multiplied by a   factor 16 as pointed out  by (*)}
 \label{Plot23basicFellParlolo} 
 \end{figure}

However, interestingly,  in Eq.(\ref{Lavalette3}),  both exponents,  among the 3-parameters, can take several signs, whence graphical forms can be quite different, as seen  in Figs. \ref{Plot23basicFellParlili}-\ref{Plot23basicFellParlolo}  shown on  these types of plots.

\begin{itemize}

\item 
but also admitting the   same  exponent $\chi$, on both tails, but  changing the range, leaving free  $N_1$ instead  of  imposing  a predetermined $(N+1)$, - of course   imposing $N_1 -r  >  0$, Ii.e.
 \begin{equation} \label{Lavalette3d}
y_N(r)= \kappa\; \big[\frac{N\;r}{ N_1-r } \big] ^{-\chi} \;\equiv\; 
\kappa\; \big[ N\;r\big] ^{-\chi} \;\big[ N_1-r \big] ^{+\chi},
\end{equation}

thus somewhat in the sense of Mandelbrot modification of Zipf law, but at high rank here.  In analogy with  the theory of critical phenomena \cite{HESbook1}, one would consider $N_1$ as the "critical range", - analogous to a "critical temperature". One  variant  of Eq.(\ref{Lavalette3d}) is merely  equivalent  to a simple redefinition of $\kappa$:   $\hat{\kappa} \equiv \kappa N^{-\chi}$. Note  again that the role of $N$ as a factor of $r$  makes "no  practical sense". Technically, for optimizing the data fits, it is better to scale the right hand side of such relations, e.g., by a factor $10^m$, $m$ obtained, in terms of  some order of magnitude of $y_N(r)$, like $y_N(N/2)$.

\item Another 3-parameter extension of Lavalette function has already been considered, - in a study of word
distribution in scientific and belletristic literature  \cite{JQL18.11.274Voloshynovska}, keeping the same exponent $\chi$ for the behavior at low and high rank, i.e.  practically replacing $r$ by some translation $r+q$ in Eq. (\ref{Lavalette2}), within a Mandelbrot trick idea. This corresponds in writing also $N_1$ for $N$ in Eq.(\ref{Lavalette3d}).
\end{itemize}

 
\section{Generalized 4-parameter free Lavalette function}\label{sec:Lav4}
The modification made in Eq.(\ref{Lavalette3d}) suggests to apply the  Mandelbrot modification also at low rank, in Eq.(\ref{Lavalette3b}),
 when there is some flattening of the data at low rank,  i.e., one introduces the a similar ZMP trick, as in Eq.(\ref{ZMeq3}) on Lavalette function,   
such that (Lav4)
 \begin{itemize}
\item
combining     Eq.(\ref{ZMeq3}) idea with the form of Eq.(\ref{Lavalette2}),  (note that it is different from Eq.(\ref{Lavalette3d})), - here keeping the same "names" for the parameters: 

\begin{eqnarray}\label{ZMeq4}
y_N(r)= \kappa\;  \frac{N^{-\chi}\;(m+r)^{-\chi}}  { (N-r+1)^{-\xi}  }
    \equiv \; 
\hat{\kappa}\; \big[m+r\big]  ^{-\chi}\; \big[N-r+1\big]  ^{+\xi} 
\end{eqnarray}
\item
another 4-parameter free generalized Lavalette function would be 
\begin{eqnarray}\label{ZMeq4mEq15}
y_N(r)= \kappa\;  \frac{N^{-\chi}\;(r)^{-\chi}}  { (N-r+m)^{-\xi}  }
    \equiv \; 
\hat{\kappa}\; \big[ r\big]  ^{-\chi}\; \big[N-r+m\big]  ^{+\xi} 
\end{eqnarray}

\item
 still a 4-parameter free generalized Lavalette function would be 
\begin{eqnarray}\label{ZMeq4m}
y_N(r)= \kappa\;  \frac{N^{-\chi}\;(m+r)^{-\chi}}  { (N-r+m)^{-\xi}  }
    \equiv \; 
\hat{\kappa}\; \big[m+r\big]  ^{-\chi}\; \big[N-r+m\big]  ^{+\xi} 
\end{eqnarray}

\end{itemize}

These differ  from  the generalization  \cite{JQL18.11.274Voloshynovska,Glottom6.03.83popescu} based on a Zipf-Mandlebrot function, because hereby allowing for different exponents $\chi$ and $\xi$.

\section{Generalized 5-parameter free Lavalette function}\label{sec:Lav5}
 "Finally", and    rather  generally a 5-parameter free function  (Lav5) is "obviously"  in order: 
\begin{equation}\label{Lav5}
y_N(r)= \kappa\;  \frac{N\;(m+r)^{-\chi}}  { (N-r+n)^{-\xi}  }
    \equiv \; 
\hat{\kappa}\; \big[m+r\big]  ^{-\chi}\; \big[N+n-r\big]  ^{-+\xi} 
\end{equation}

No graph  illustrates  this super-generalization; a simple combinatory calculation indicates that one would ask for ten of them. It is better to  suggest to envisage such a form when those with  a lower number of free parameters do  not lead to satisfactory or successful fits. It seems that one can rather easily understand the effect of the new parameters when examining the functions.

\section{  A few other  formulae for fits}\label{sec:afewformulae}
  
  For completeness, recall a few other often used formulae for fitting data (often)  on log-log plots.
\subsection{2 parameters}\label{sec:2parameters}
 \vskip0.5cm
 
 Beside the power law, Eq.(\ref {Zipfeq})  and    the  basic  2-parameter Lavalette  form,   Eq.(\ref{Lavalette2}), 
 one should mention 
  \begin{itemize}
 
 \item
 the   (2 parameter) exponential case  \begin{equation} \label{EXP2}
 y(r)=  \;b\;e^{-\beta r}
\end{equation}  
\item  
a law  suggested by Tsallis and  de Albuquerque\footnote{correcting a misprint in \cite{Glottom6.03.83popescu}. } (for ranking paper citations)  \cite{TdA}  
 \begin{equation} \label{TdA2}
y(r)=\frac{\phi}{[1+(\psi'-1]\; ln(r))^{\psi}}
\end{equation}
with  $\psi' \equiv \psi$, although  there does not seem any reason why it should be so.

\item
the   log-normal distribution  \cite{[11]},  

\begin{equation} \label{lognormaleq}
y(x) = \frac{1}{x \sigma \sqrt{2\pi}}\; exp(-\frac{(lnx -\mu)^2}{2\sigma^2})
\end{equation}
 where $x>0$, $\mu$ and $\sigma$ are the parameters, mean and standard deviation of the log of "variable" in  the data distribution.
\end{itemize}

\subsection{3 parameters}\label{sec:3parameters}

Beside  
  the  Verhulst logistic form, Eq.(\ref{Vlog3}) and  the  Zipf-Mandelbrot-Pareto  (ZMP3) law  \cite{FAIRTHORNE}, Eq.(\ref{ZMeq3}), other  often used  3-parameter  statistical distributions,  generalizing the  power  and/or exponential law   are to be examined :  
 \begin{itemize}
 \item 
 the  Yule-Simon distribution, i.e. a power law with exponential cut-off  \cite{Pwco3}   (the free parameters are: $d$, $\alpha$, and $\lambda$)
  \begin{equation} \label{PWLwithcutoff}
 y(r)= d \;r^{-\alpha} \; e^{-\lambda r},
\end{equation}

 \item
   the stretched exponential \cite{Stx3}    (the free parameters are: $\theta$, $\mu$, and $\nu$)
 \begin{equation} \label{Stxeq3}
y(r)=\theta \; x^{\mu-1} \; e^{-\nu\;x^\mu}.
\end{equation}
\item   
the Gompertz double exponential  \cite{G2xp}    (the free parameters are: $g_1$, $r_2$, and $ g_3$)
\begin{equation}
\label{Glowt}
y(r)= g_1\;e^{-e^{-(r-r_2)/g_3}}
\end{equation}



\end{itemize}

These    function also bend  in 	 convex form on a log-log plot.
 
\subsection{4 parameters}\label{sec:4parameters}

There are  several possible generalizations of the above, often introducing the Mandelbrot trick, at low rank,  i.e. $r\; \rightarrow\; r+ \rho$, with a possibly different $\rho$ at high  and low ranks, but they do not seem of major interest.   Indeed, look at  

\begin{itemize}
\item
a  ZMP4 form, e.g., 

\begin{equation}\label{ZMeq4b}
y(r)\; =\; m_3 /(m_2+ m_4\;r)]^{\zeta},
\end{equation}
which obviously reduces to  Eq.(\ref{ZMeq3}) by a trivial change in the parameter notations, e.g. $\hat{m_3} \rightarrow m_3/m_4^{\zeta}\equiv\; c$, and $m_2/m_4 \equiv \eta$,
\item 
or 
 \begin{equation} \label{MApwlco4}
 y(r)=m_3 \; (r-m_4)^{- m_1} \; e^{ -m_2\;(r-m_4)  }
\end{equation}
with $m_4\equiv$ to some $r_0$,  which  it is nothing else that
 \begin{equation} \label{MApwlco4b}
 y(r)=\hat{m_3} \; (r-m_4)^{- m_1} \; e^{ -m_2\;r   }.
\end{equation}
\end{itemize}

Usually  such functions reproduce one   tail  but not the other.  

Technically, such  improvements do not change in a dramatic way the  regression  coefficient,   since the  high rank tail  does not have a great impact upon this coefficient, - because of the change in the order of magnitude between the low and high rank regions.

\section{Hyper-generalized (Lavalette) fit functions} \label{sec:HyperLav}
 
   \begin{figure}
\centering
 \includegraphics
 [height=7.4cm,width=11.0cm]{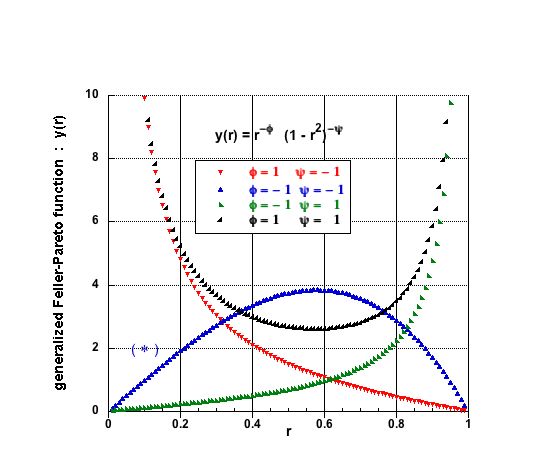}
\caption   {  Hyper-generalized  Feller-Pareto function, $y(r) = r^{\phi}   (1 - r^2)^{-\psi}$, on ordinary axes;  (*) indicates that the function has been multiplied by a factor 16  for better readability} 
 \label{Plot5BCDE10C4gFPLlili} 
 \end{figure}
   \begin{figure}
\centering
 \includegraphics [height=7.4cm,width=11.0cm]{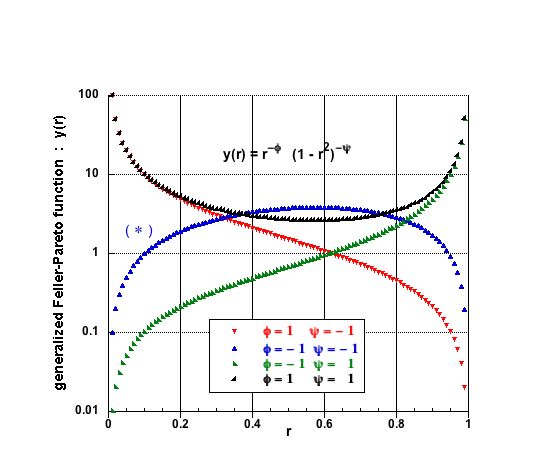}
\caption   { Hyper-generalized  Feller-Pareto function,$y(r) = r^{\phi}   (1 - r^2)^{-\psi}$, on a semi-log plot;  (*) indicates that the function has been multiplied by a factor 16  for better readability} 
 \label{Plot5BCDE10C4gFPlilo} 
 \end{figure}
    \begin{figure}
\centering
 \includegraphics [height=7.4cm,width=11.0cm]{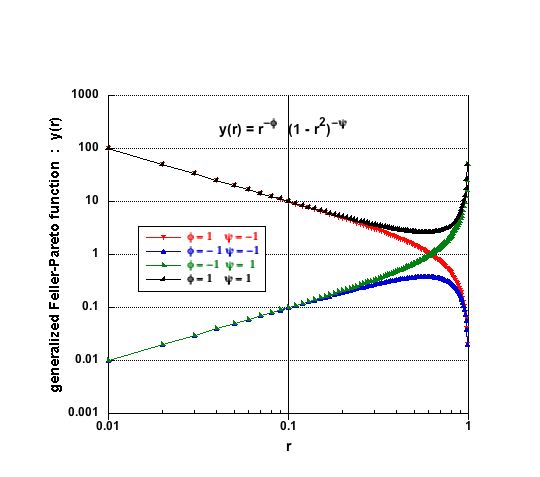}
\caption   { Hyper-generalized Feller-Pareto function, $y(r) = r^{\phi}   (1 - r^2)^{-\psi}$, on a log-log plot
} 
 \label{Plot10BCDElolo4FP} 
 \end{figure}

It might be reminded that the modification of Keynes  differential growth equation by Verhulst through a ($1-x$) term was purely a  mathematical  {\it ad hoc} mean to avoid a full exponential growth. There is no economic or demographic argument to use a linear ($1-x$) term; a quadratic term ($1-x^2$) or any other polynomial decaying near $x=1$ or many more complicated terms could be used.  Therefore,considering that the basic phenomena might not necessarily depend linearly on $r$, but the rank-size rule should  (or could) contain higher order terms,  other generalizations may come  in mind  within the present considerations.  One such a case was found in considering city sizes (in Bulgaria, e.g. \cite{ZDMA,NKVZDMA}), but might occur more frequently than "expected", - however are  likely not reported because of missing framework. Therefore, hyper-generalizations of Lavalette function can be imagined:

 \begin{itemize}
\item   the  3-parameter generalized Lavalette form   \cite{JoI1.07.155Mansilla}  can be hyper-generalized into
 \begin{equation} \label{Lavalette3amn}
y(r)=   \frac{\big[\Lambda \;r^n\big]^{-\phi}} { \big[N+1-r^m\big] ^{-\psi} }
 \;\;    or  \;\;   =  \Lambda \;\frac{\big[r^n\big]^{-\phi}} { \big[N+1-r^m \big] ^{-\psi} }
\end{equation}

\item  the  4-parameter generalized Lavalette form   \cite{JQL18.11.274Voloshynovska}  can be hyper-generalized into
 \begin{equation} \label{Lavalette4mn}
y(r )=   \frac{\big[\Gamma \;(r^n+\nu)\big]^{-\eta}} { \big[N-r^m+\nu\big] ^{-\zeta} } 
 \;\;    or  \;\;   =\Gamma \frac{\big[r^n+\nu\big]^{-\eta}} { \big[N-r^m+\nu\big] ^{-\zeta} } 
\end{equation}

\item    the  5-parameter  supergeneralized Lavalette form    (also)  can be hyper-generalized into
 \begin{equation} \label{Lavalette5mn}
y(r )=   \frac{\big[\Gamma \;(r^n+\mu)\big]^{-\eta}} { \big[N-r^m+\nu\big] ^{-\zeta} } \;\; or  \;\;  = \Gamma \;
\frac{\big[(r^n+\mu)\big]^{-\eta}} { \big[N-r^m+\nu\big] ^{-\zeta} } 
\end{equation}

\end{itemize}

Note that variants : $ \big[   (r^n+\nu)\big] \rightarrow \big[ (r+\nu)^n\big]$,  and  $ \big[  (r^n+\mu)\big] \rightarrow \big[ (r+\mu)^n\big]$,  with or without  $ \big[  \;(r^m-\nu)\big] \rightarrow \big[ (r-\nu)^m\big]$, can be written, connecting to the (2-parameter free) Burr function \cite{Burr49}. The writing choice is left for fit optimization time.


\section{On inflection points on log-log plots} \label{sec:inflectionloglog}

Finally, not not the least,  the above formulae have much emphasized possible  fits which indeed allow inflection points on semi-log graphs, but have left opened the case of inflection points on log-log graphs. Let it be understood that such a case occurs  when some poser law decay  ("from infinity") at low rank is followed by another  intermediary regime before  some cut-off occurs at high rank. A trivial transformation $x \rightarrow log(x)$ of all the above formulae is possible, but  demands much reflection. Indeed, one could transform the basic Lavalette equation to read

\begin{equation}\label{xlogr}
y(r) \; \simeq \; \big[\frac{N\; log (r)}{N+1-log(r)}\big]^{-\chi}
\end{equation}
and similarly all others. But it remains some interpretation and  much theoretical work !

Another possibility comes from realizing that if there is an inflection point, the slope has the same  (negative) sign  for the whole $r$ range, but the derivative of the slope has some  structure, i.e.allowing for a concave  to a convex shape of the approximation  to the data. The intermediary regime can also be considered in a first approximation to be a scaling law. The high rank regime can be either a Lavalette cut-off or an exponential cut-off. Therefore the following functions can be  appropriately imagined
\begin{itemize}
\item in its most generalized form, with  power law cut-off
\begin{equation}\label{m1m7Lav}
y(x) = \big[A \;(x+m_5)^{-m_1} + B\; (x+m_6)^{-m_2}\big]\; (N+m_4-x^{m_7})^{m_3}
\end{equation}

\item  or with an exponential cut-off
\begin{equation}\label{m1m7exp}
y(x) = \big[A \;(x+m_5)^{-m_1} + B\; (x+m_6)^{-m_2}\big]\; e^{-m3\;(x+m4^{m7}}
\end{equation}
\end{itemize}

A few of such cases are shown in Figs. \ref{Plot13(c0(-m) +c0(-n))(N-c0)copy}-\ref{Plot7lastfig} 
demonstrating the interest of such forms in order to discuss  inflection points on log-log plots.

 \section{Applications}\label{sec:applications}
 
 This section serves as an illustration of a few   cases discussed above,   displaying some data on either semi-log or log-log plots for comparison. However the data pertains to some empirical study requesting a brief introduction. In so doing, it is hoped that the "universality" of the approach receives a positive argument. 
 
 Consider the following investigation. In Italy, 638 cities contain a saint  or an angel name, as counted after translating the names into italian, from french, german, or local dialects (like Santu Lussurgiu  =  Santo Lussorio, or Santhi\' a who is Santa Agata), Note that   Sant'Angelo (24 times), San Salvatore (5 times) or Santa Croce  (7 cities), and similar  "concepts"  (Sansepolcro) are not counted. Some distinction can be made between male and female saints. Note that two cities have a name with two saints. The name of the saints can be ranked according to  their frequency \cite{MARCITsaints} and an appropriate statistical analysis can follow  for the rank-frequency distribution. 
 
 However,  one can also ask, as did Pareto in 1896,  how many times one can find 
  an  "event" greater than some size $n$, i.e. study  the {\it size-frequency  relationship}. Pareto  found out that  the   cumulative distribution function (CDF) of such events  follows an inverse power of $n$, or  in other words,  $P\;[N>n] \sim n^{-\omega}$., - whence the frequency $f$ of such events   of size $n$, (also) follows an inverse power of $n$. 
 
\begin{figure}
\centering
 \includegraphics [height=14cm,width=14.0cm]{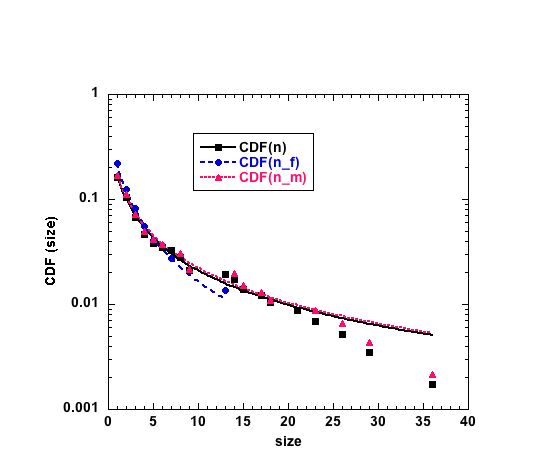}
\caption   {\underline{Semi-log} plot of the cumulative distribution function  (CDF) of the frequency of  Italian cities containing a saint name $n$-times, so called "size",  given according to the Zipf-Mandelbrot-Pareto function,  like Eq.(\ref{ZMeq3}),   distinguishing between male (n$_{-}$m) and female  (n$_{-}$f)  saint names; the fit parameter values are given in Fig.\ref{Plot42CDFloloZMP3}. 
Observe the need for a cut-off at high rank/size
} 
 \label{Plot42CDFliloZMP3} 
 \end{figure}

\begin{figure}
\centering
 \includegraphics [height=14cm,width=14.0cm]{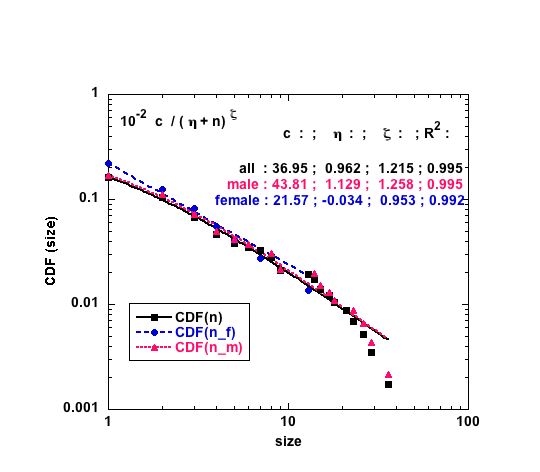}
\caption   {Log-log plot of the cumulative distribution function  (CDF) of the frequency of  Italian cities containing a saint name $n$-times given according to the Zipf-Mandelbrot-Pareto function, like Eq.(\ref{ZMeq3}),  distinguishing between male (n$_{-}$m) and female  (n$_{-}$f) saint names;  observe that $\eta$  is negative for the female case, pointing to a king effect (Santa Maria), and queen effects, since $\eta \ge0$, for the males and the overall distribution. 
Observe the need for a cut-off at high rank/size}
 \label{Plot42CDFloloZMP3} 
 \end{figure}

\begin{figure}
\centering
 \includegraphics [height=14cm,width=14.0cm]{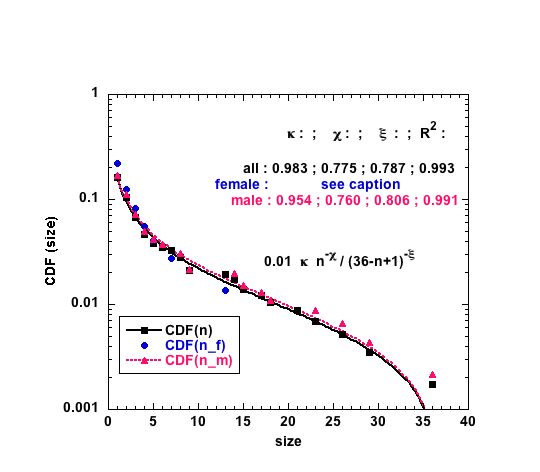}
\caption   { \underline{Semi--log} plot of the cumulative distribution function  (CDF) of the frequency of  Italian cities containing a saint name,  given $n$-times, so called "size";   fit according to a Lavalette  function with 3 free parameters, Eq.(10), for  the distribution of all such 36 cities(black line)  or only those 36 with a male saint name (n$_{-}$m; red line); the parameter values for the female case are given in Fig.\ref{Plot42CDFloloLav3}, with the  corresponding fit. Observe the interest of leaving the   high rank/size value be a free parameter, as on Fig.\ref{Plot42CDFliloLav4Eq15} 
} 
 \label{Plot42CDFliloLav3} 
 \end{figure}

\begin{figure}
\centering
 \includegraphics [height=14cm,width=14.0cm]{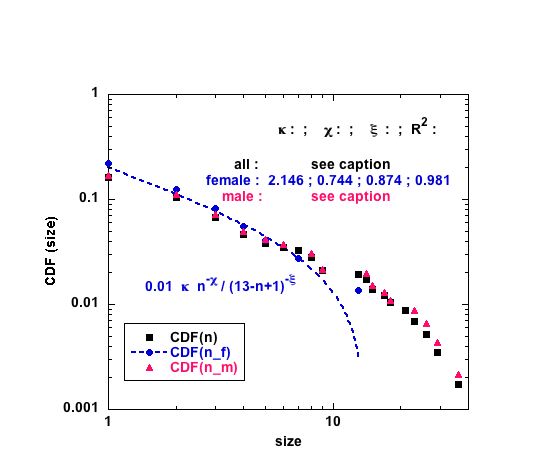}
\caption{ Log--log plot of the cumulative distribution function  (CDF) of the frequency of  Italian cities containing a saint name $n$-times  given $n$-times, so called "size";   fit according to a Lavalette  function with 3 free parameters, Eq.(10) is shown for   the distribution of  only those 13 cities with a female saint name (n$_{-}$f; blue line); the parameter values for the male case and the whole distribution are given in Fig.\ref{Plot42CDFliloLav3}, with the  corresponding fits. Observe the interest of leaving the   high rank/size value be a free parameter, as on Fig.\ref{Plot42CDFloloLav4Eq15} 
} 
 \label{Plot42CDFloloLav3} 
 \end{figure}

\begin{figure}
\centering
 \includegraphics [height=14cm,width=14.0cm]{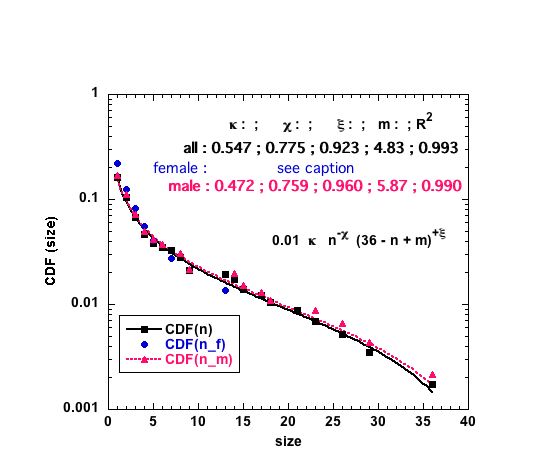}
\caption   { \underline{Semi--log }plot of the cumulative distribution function  (CDF) of the frequency of  Italian cities containing a saint name  given $n$-times, so called "size"; fits with a 4 parameter free Lavalette  function,  Eq.(\ref{ZMeq4mEq15})  are shown for  the distribution of all such 36 cities (black line)  or only those 36 with a male saint name (n$_{-}$m; red line); the parameter values for the female case are given in Fig.\ref{Plot42CDFloloLav4Eq15}, with the  corresponding fit   } 
 \label{Plot42CDFliloLav4Eq15} 
 \end{figure}

\begin{figure}
\centering
 \includegraphics [height=14cm,width=14.0cm]{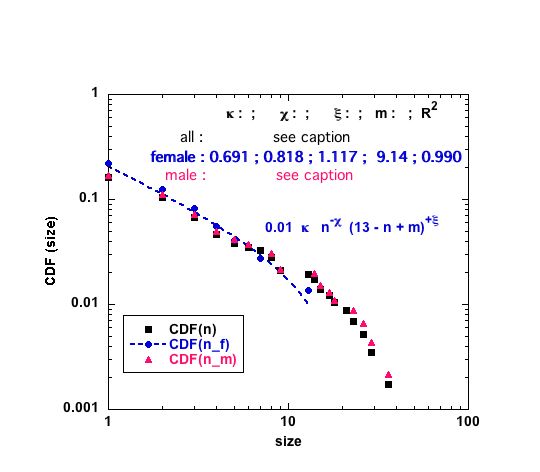}
\caption   { Log-log plot of the cumulative distribution function  (CDF) of the frequency of  Italian cities containing a saint name  given $n$-times, so called "size";   fit according to a Lavalette  function with 4 free parameters,  Eq.(\ref{ZMeq4mEq15}) shown for   the distribution of  only those 13 cities with a female saint name (n$_{-}$f; blue line); the parameter values for the male case and the whole distribution are given in Fig.\ref{Plot42CDFliloLav4Eq15}, with the  corresponding fits
} 
 \label{Plot42CDFloloLav4Eq15} 
 \end{figure}

\begin{figure}
 \centering
 \includegraphics 
  [height=14cm,width=14cm] 
  {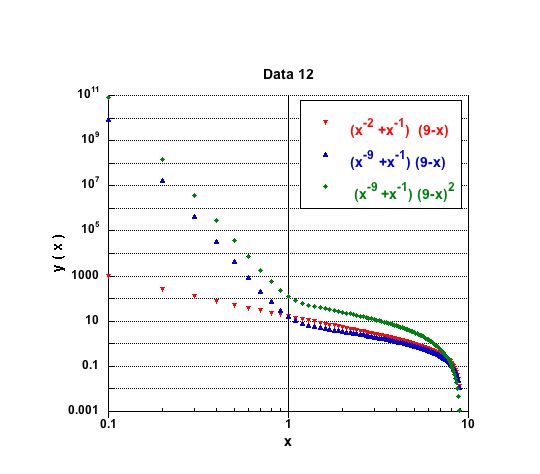}
\caption   { Display of a "simple" function with inflection point on a  log-log plot, allowing for fit to data with large king or queen effect and power law cut-off, i.e. with an inflection point in the middle range, as approximated by a simple function for which the general form is Eq.(\ref{m1m7Lav})
 } 
 \label{Plot13(c0(-m) +c0(-n))(N-c0)copy} 
 \end{figure}

\begin{figure}
 \includegraphics
  [height=14cm,width=14cm] {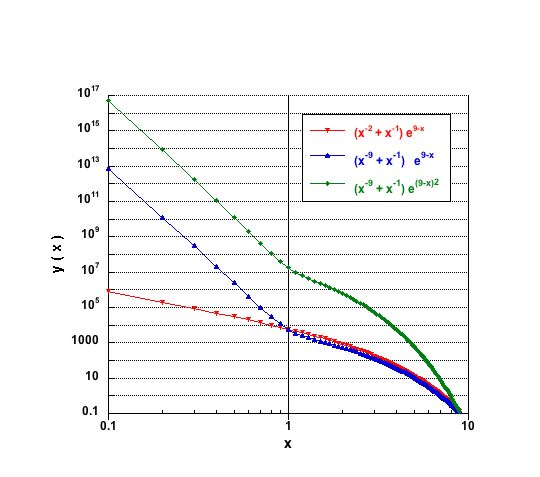}
\caption   { 
Display of a "simple" function with inflection point on a  log-log plot, allowing for fit to data with  large king or queen effect and  exponential  cut-off, i.e. with an inflection point in the middle range, as approximated by a simple function for which the general form is Eq.(\ref{m1m7exp})
 } 
 \label{Plot7lastfig} 
 \end{figure}

Thus, one can count how many  cities have a happax hagionym, how many cities have a  name with a saint occurring only twice, etc.  up to how many cities have a name associated to the "most popular"  (=  most frequent)  saint ( San Pietro). This counting is normalized and turned into a probability distribution, i.e. CDF(n).  The data is illustrated in Figs. \ref{Plot42CDFliloZMP3}-\ref{Plot42CDFloloLav4Eq15}, either with semi-log or log-log plots, and fits  with a Zipf-Mandelbrot  or Lavalette function.
 
Short final comments:  (i) two   "queen effects"   
     and   a  "king effect"    are  well seen  on Fig. \ref{Plot42CDFloloZMP3};  (ii) the CDF shows a pronounced cut-off at high $n$ in all cases. Therefore,  it could  be argued that the CDF is less pertinent to observe minute effects. This is understandably true, since the CDF results from an integration scheme. However, again understandably, the CDF fits are much more stable. No need to say that one should not report too precise parameter values, since these are non linear fits;  a final technical information: the Levenberg-Marquardt algorithm was used.
    
 
 
\section{Conclusions}\label{sec:conclusions}

It has been shown that semi-log plots are of interest in order to analyze whether 
experimental or  empirical data  are underlined by some scaling argument for the observed/examined phenomenon at hands.  The fit to a straight line on log-log plots is not always  satisfactory indeed. Deviations occur at low, intermediate and high regimes along the  $x$-axis. Several improvements of the mere power law fit  have been discussed, in particular through a Mandelbrot trick at low rank and a Lavalette power law cut-off at high rank. 

In so doing \cite{JoI1.07.155Mansilla,JQL18.11.274Voloshynovska,Glottom6.03.83popescu}, the number of free parameters increases. Their meaning has been discussed, up to the  5 parameter free super-generalized  Lavalette law and the 7-parameter free hyper-generealized Lavalette law\footnote{In this conclusion, one could recall that  7-parameter free functions are also used for fitting data like in financial market crash predictions  \cite{SornetteAJBouchaudJFI96,how,sornetteNASDAQ,bigtokyo,DrozdzPhRep515} or in earthquake predictions \cite{kobe} } . It  has been emphasized that the interest of the  basic 2-parameter free Lavalette law and the subsequent generalizations resides in its "noid" (or sigmoid, depending on the sign of the exponents) form on a semi-log plot; something incapable to be found in other empirical law, like the Zipf-Pareto-Mandelbrot law.  The connection with other laws, e.g. Feller-Pareto and Verhulst logistic laws, have been pointed out. 

It  has been shown that 
the additional   parameters introduced into the  basic Lavalette 
function, Eq.(\ref{Lavalette2}), facilitates a rather good reproduction of rank-probability
distribution in the ranges  of small and high rank values. Indeed, each parameter or ratio involved in the suggested modification of Lavalette function, Eq.(\ref{Lavalette2}), enhances the fit in different ranges of  $r$.  

It has remained for completeness to invent a simple law showing an inflection point on a \underline{log-log plot}. Such a law could have been the  result  of a transformation of the Lavalette law through $x$ $\rightarrow$  log($x$), but this meaning is theoretically unclear.  It has been shown that a simple linear combination of two basic Lavalette law is  provides the requested features. Generalizations taking into account two super-generalized or hyper-generalized Lavalette laws are suggested, but need to be fully considered at fit time on appropriate data.
 
A few examples are used for illustrating various points, like deviations or visually unattractive fits, - though the regression coefficient $R^2$ is often quite satisfactory looking. Examples have been taken mainly  for rank-size rule research. However, in order to demonstrate a larger validity of  generalizing the usual  fit formulae, and some interest for generalizing the basic concepts,  some  short analysis  has been presented of  the   cumulative distribution function (CDF) of the city names in Italy containing a  (male or female) saint name.

  \bigskip  \bigskip 

{\bf Acknowledgements} Thanks to  C. Herteliu for  comments prior to manuscript submission.  I thank all colleagues  mentioned in the text and bibliography for  providing relevant data and any sharp remark. This paper is part of scientific activities in COST Action TD1210.

This paper is part of scientific activities in COST Action COST
Action IS1104, "The EU in the new complex geography of economic
systems: models, tools and policy evaluation".
 \bigskip


\end{document}